\documentclass[prl,twocolumn,numbers,amsmath]{revtex4}
\usepackage{graphicx}
\usepackage{dcolumn}
\usepackage{bm}
\usepackage{epsfig}
\usepackage{wrapfig}

\begin{document}

\title{Absence of halfmetallicity in defect-free Cr, Mn-delta-doped \\ Digital Magnetic Heterostructures}
\author{F. Beiu\c seanu, %\footnote{Corresponding author e-mail:fbeiu@uoradea.ro}
C. Horea, E.-V. Macocian, T. Jurcu\c t}
\affiliation{Department of Physics, University of Oradea, RO-410087 Oradea, Romania}
\author{L. Vitos}
\affiliation{Applied Materials Physics, Department of Materials Science and Engineering,
Royal Institute of Technology, Stockholm SE-100 44, Sweden}
\affiliation{Department of Physics and Materials Science, Uppsala University, P.O. Box 516, SE-75120 Uppsala, Sweden}
\affiliation{Research Institute for Solid State Physics and Optics, Budapest H-1525, P.O. Box 49, Hungary}
\author{L. Chioncel}
\affiliation{Augsburg Center for Innovative Technologies, University of Augsburg, D-86135 Augsburg, Germany}
\affiliation{Theoretical Physics III, Center for Electronic Correlations and Magnetism, Institute of Physics,
University of Augsburg, D-86135 Augsburg, Germany}

\begin{abstract}
We present results of a combined density functional and 
many-body calculations for the electronic and magnetic properties 
of the defect-free digital ferromagnetic heterostructures obtained 
by %delta-
doping GaAs with Cr and Mn. 
While local density approximation/(+U) predicts half-metallicity in these 
defect-free delta-doped heterostructures, we demonstrate that local  
many-body correlations captured by Dynamical Mean Field Theory induce 
within the minority spin channel non-quasiparticle states just above $E_F$.
As a consequence of the existence of these many-body states the half-metallic 
gap is closed and the carriers spin polarization is significantly reduced. 
Below the Fermi level the minority spin highest valence states are found 
to localize more on the GaAs layers being independent of the type of 
electronic correlations considered. Thus, our results confirm the confinement 
of carriers in these delta-doped heterostructures, having a spin-polarization 
that follow a different temperature dependence than magnetization. We suggest 
that polarized hot-electron photoluminescence experiments might bring evidence 
for the existence of many-body states within the minority spin channel and their
finite temperature behavior.
\end{abstract}
\maketitle

\section{Introduction}
Digital Magnetic Heterostructures (DMH) are semiconductor heterostructures 
in which magnetic mono-layers are incorporated using the digital-alloy 
technique \cite{cr.tu.95,aw.sa.99}. This technique was developed in order to combine the 
conventional semiconductors with magnetic materials, with the desire to 
realize simultaneous band gap and magnetic engineering. Within the DMH 
technique, the quality of magnetic mono-layers is highly sensitive to 
growth conditions similarly to the case of diluted magnetic semiconductors,
were low temperature thermal annealing \cite{ed.bo.04,li.ko.05} and 
codoping \cite{pa.li.03,ju.ma.03} has been extensively developed. The existing 
technological difficulties increase therefore the significance of 
first-principles investigation, which can predict many characteristics of 
DMH and indicate trends in their properties.

For DMH materials, most first-principles investigations were done for 
GaAs/Mn \cite{sanv.03,xu.sc.06}. The exchange interactions inside and between 
Mn mono-layers were analyzed, in particular the spacer dependent magnetic
properties were recently discussed \cite{wa.ch.09}. Independent of the
spacer thickness the electronic structure was predicted to be half-metallic and using
the technique of injection of free holes \cite{qi.fo.06b} intra and interlayer exchange
coupling mechanisms were discussed \cite{wa.ch.09}.
In addition, previous studies  discussed transport properties, optimal concentration of Mn, 
possible effect of anti-site defects, etc., \cite{sanv.03,xu.sc.06} . A number of 
calculations were also performed for Mn doped Si, Ge, and Ga \cite{qi.fo.06,wu.kr.07,ma.fe.06}.

Properties of these heterostructures are expected to deviate from the 
physical behavior presented by family of diluted magnetic semiconductors \cite{ju.si.06}. 
The most common position occupied by Mn within the (III,V) host GaAs diluted magnetic 
semiconductor is the Ga site. The substitutional Mn contributes with its two 4s electrons 
in the crystal bonding similarly as the two Ga-4s electrons. Less favorable energetically
is the interstitial Mn position and therefore is less common \cite{ju.si.06}. 
While in Mn doped GaAs semiconductors the system is described as ``bulk", in 
the fabrication of GaAs/Mn-digital magnetic heterostructures, MnAs 
mono-layers were embedded into GaAs superlattices \cite{ka.jo.00}, so that the 
quasi-bidimensional geometry of the MnAs layer is preserved. Thus, the 
random substitution of positions (Ga occupied or interstitial) by the Mn 
ions in the case of diluted magnetic semiconductors shows up as 
an ordered replacement of a GaAs layers by MnAs layers in the case of DMH, 
with evident consequences upon the critical temperature. DMH exhibits an 
unexpected Curie temperature dependence on concentration of Mn ions. At a 
fixed value of Mn concentration (fixed number of MnAs layers), the Curie 
temperature decrease with the thickness of the GaAs, but saturates at large 
thicknesses ($\cong 50$ GaAs mono-layers)\cite{ka.jo.00}. 
Electronic structure calculations were also reported in the literature for the 
case of $\delta$-doping with Mn of GaAs/AlAs superlatices \cite{sanv.03}. In 
the defect free cases a half-metallic heterostructure was obtained. It 
was also demonstrated that in the presence of As-anti-sites, the half-metallic
state was destroyed and a {\it p}-type metallic conductance was evidenced
because of majority spin electrons. In addition the computed magnetic 
coupling between the Mn ions becomes larger in the presence of 
As anti-sites \cite{sanv.03}. To our best knowledge no results which include 
electronic correlations were reported for the case of Cr or Mn even in the 
defect-free cases. 

In this paper we present results of the first-principles electronic 
structure calculations within Local Spin Density Approximation (LSDA), 
LSDA+U and LSDA+Dynamical Mean Field Theory 
(DMFT) for the different defect free (MAs)$_1$/(GaAs)$_7$ digital 
heterostructures, containing transition metal-arsenide mono-layers 
(M=Cr,Mn) introduced in the GaAs superstructure. 
The most important correlation effect in half-metals, the
formation of many-body states within the gap can not be captured by the mean-field
LSDA/+U-type calculations. Dynamical mean-field theory was demonstrated to be
able to describe such effects \cite{ka.ir.08}, therefore is essential to compare directly the 
mean-field LSDA/+U and the DMFT results. 
We demonstrate that the half-metallic character is lost in 
$\delta$-doped GaAs heterostrcutures containing Cr or Mn at finite 
temperatures and in the presence of dynamic correlations.
Correlation effects soften the magnetic properties of the ferromagnetic layers, 
induced states above $E_F$ with tails that cross the Fermi level and reduce the 
carrier spin polarization.  Within the minority spin channel below the Fermi 
level the character of states forming the band-edge remains not affected by 
electronic correlations and the states are mostly localized on GaAs layers 
confirming the experimentally observed confinement of 
carriers in these heterostructures~\cite{ka.jo.00,sa.tr.08}. 

\section{Crystal structure and computational method}

To compute the electronic structure for the digital magnetic 
heterostructures we used the standard representation of the tetragonal 
symmetry, space group $P-4m2$. The tetragonal super cell 
was obtained from the FCC unit cell by a rotation in the basal plane and an 
integer translation along the (001) direction. The new translation vectors 
are: $a = a_{0} / \sqrt{2}$, $b = a_{0} / \sqrt{2}$, $c = a_{0} N$, 
where $a_{0}$ is the bulk GaAs lattice parameter, and $N=4$, and they 
correspond to a (GaAs)$_{\rm 8}$ super cell. Within this super cell a GaAs 
mono-layer is replaced by MAs, such that the obtained DMH has the unit cell 
formula (MAs)$_{\rm 1}$/(GaAs)$_{{\rm 2}{\rm N}{\rm -}{\rm 1}}$.
For GaAs which has an open structure, a close 
packed structure is obtained by including empty-spheres. Such empty potential wells were used also 
in the present heterostructure geometry. We used identical muffin-tin spheres for all 
atoms in the unit cell having the average Wigner-Seitz radius $2.62852% 
a.u.$ corresponding to the experimental GaAs lattice parameter ($a_0=5.65\AA$) . The basis used 
for the self-consistent calculations contains the {\it spd}-partial waves for 
transition metals, $sp(d)$-partial waves for Ga, As and $s(p)$-partial waves for 
the empty spheres E. $(l)$ means that the $l$-partial waves are downfolded 
within the self consistent calculations. Selfconsistency was performed for 
105 irreducible $k$-points and total energy convergence was achieved with an 
accuracy of $10^{{\rm -} {\rm 6}}$ Ry. Results for the density of states, 
magnetic moments and spin polarizations are presented in the next section. 

Correlation effects in the valence $3d$ 
orbitals are included via an on-site electron-electron interaction in the form
$\frac{1}{2}\sum_{{i \{m, \sigma \} }} U_{mm'm''m'''}
c^{\dag}_{im\sigma}c^{\dag}_{im'\sigma'}c_{im'''\sigma'}c_{im''\sigma} $.
The interaction is treated in the framework of DMFT
\cite{ko.vo.04,ko.sa.06,held.07}, with a spin-polarized T-matrix Fluctuation Exchange (SPTF) type
of impurity solver \cite{ka.li.02} implemented within the EMTO basis set
\cite{vito.01,ch.vi.03}. 
The Coulomb matrix elements $U_{mm'm''m'''}$ are expressed in the usual
way~\cite{im.fu.98} in terms of  three Kanamori parameters $U$, $U'=U-2J$
and $J$.
Here, $c_{im\sigma}/c^\dagger_{im\sigma}$
destroys/creates an electron with spin $\sigma$ on orbital $m$ on
lattice site $i$.
The SPTF approximation is a multiband
spin-polarized generalization of the fluctuation exchange
approximation (FLEX) \cite{bi.sc.89,ka.li.99}, but with a
different treatment of particle-hole (PH) and particle-particle
(PP) channels. The particle-particle (PP) channel is described by
a $T$-matrix approach \cite{gali.58,kana.63} giving a renormalization
of the effective interaction. This effective interaction is used
explicitly in the particle-hole channel. Justifications, further
developments and details of this scheme can be found in Ref.
\onlinecite{ka.li.02}.

For the case of half-metallic ferromagnets it was demonstrated
\cite{ka.ir.08} by model as well as realistic electronic structure calculations that many-body
effects are crucial for half-metals: they produce states with tails that cross the Fermi level
so that the gap is closed and half-metallicity is lost~\cite{ch.ka.03,al.ch.10,ch.ma.06,ch.ka.05,ch.ar.09}.
The origin of these many-body non-quasiparticle (NQP) states is connected with ``spin-polaron'' processes:
the spin-down low-energy electron excitations, which are forbidden for the HMF in the
one-particle picture, turn out to be possible as superpositions of
spin-up electron excitations and virtual magnons \cite{ed.he.73,ka.ir.08}.
Spin-polaron processes are described within the SPTF approach
by the fluctuation potential matrix $W^{\sigma \sigma ^{\prime
}}(i\omega )$ with $\sigma=\pm$, defined in a similar way as in
the spin-polarized FLEX approximation \cite{ka.li.99}:

\begin{equation}\label{W}
{\hat W}(i\omega )=\left(
\begin{array}{cc}
{W}^{++}(i\omega ) & {W}^{+-}(i\omega ) \\
{W}^{-+}(i\omega ) & {W}^{--}(i\omega )
\end{array}
\right).
\end{equation}

The essential feature here is that potential (\ref{W}) is a
complex energy dependent matrix in spin pace with {\it
off-diagonal} elements:
\begin{equation}
W^{\sigma, -\sigma}(i\omega)=U^m(\chi^{\sigma, -\sigma}(i\omega )-
\chi_0^{\sigma, -\sigma}(i\omega ))U^m
\end{equation}
where $U^m$ represents the bare vertex matrix corresponding to the
transverse magnetic channel, $\chi^{\sigma, -\sigma}(i\omega )$ is
an effective transverse susceptibility matrix and $\chi^{\sigma,
-\sigma}_0(i\omega )$ is the bare transverse susceptibility
\cite{ka.li.99}. $i\omega$ are fermionic Matsubara frequencies
and $(m)$ corresponds to the magnetic interaction channel
\cite{bi.sc.89,ka.li.99}. The local Green functions as well as
the electronic self-energies are spin diagonal for collinear
magnetic configurations. In this approximation the electronic
self-energy is calculated in terms of the effective interactions
in various channels. The particle-particle contribution to the
self-energy  was combined with the Hartree-Fock and the second
order contributions \cite{ka.li.99}. To ensure a  physical
transparent description the combined
particle-particle self-energy is presented by its a Hartree
$\Sigma^{(TH)}(i\omega )$ and Fock $\Sigma^{(TF)}(i\omega )$ types of contributions:
%\begin{equation}
$\Sigma(i\omega ) = \Sigma^{(TH)}(i\omega )+ \Sigma^{(TF)}(i\omega ) + \Sigma^{(ph)}(i\omega )$
%\end{equation}
where the particle-hole contribution $\Sigma^{(ph)}$ reads:
\begin{equation}\label{selfph}
\Sigma_{12 \sigma} ^{(ph)}(i\omega ) = \sum_{34 \sigma^{\prime}} W_{1342}^{\sigma
\sigma^{\prime}}(i\omega ) G_{34}^{\sigma^{\prime}}(i\omega )
\end{equation}
Since the static contribution from correlations is already
included in the local spin-density approximation (LSDA), so-called
``double counted'' terms must be subtracted. In other words,
those parts of the DFT expression for the total energy
that correspond to the interaction included in the Hubbard
Hamiltonian has to be subtracted.
Several double-counting schemes 
has been proposed and used for specific materials applications \cite{an.ar.97,ya.sa.01}. It has been 
recognized that the results of the LSDA+U calculations may depend crucially
on the choice of the scheme used~\cite{pe.ma.03}.  In particular for the class of moderately 
correlated metals as one may consider the half-metallic ferromagnets in  the 
LSDA+U calculations performed here we used the ``around mean-field''~\cite{pe.ma.03}
double-counting scheme. Similarly to the LSDA+U in the DMFT approach the Hubbard
Hamiltonian represents the underlying physics, while in contrary to the mean-field
LSDA+U solution the time dependent dynamics are considered, while spatial 
fluctuations are neglected. Obviously double counting correction has to be 
considered also. To achieve this, we replace
$\Sigma_{\sigma}(E)$ with $\Sigma_{\sigma}(E)-\Sigma_{\sigma}(0)$
\cite{li.ka.01} in all equations of the DMFT procedure \cite{ko.sa.06}.
Physically, this is related to the fact that DMFT only adds {\it dynamical}
correlations to the LSDA result. For this reason, it is believed that this kind
of double-counting subtraction ``$\Sigma(0)$'' is more appropriate for a DMFT
treatment of metals than the alternative static ``Hartree-Fock'' (HF) subtraction
\cite{pe.ma.03}.

\section{Results and discussion}
When the mono-layer of MAs, with M=Cr, Mn is
introduced in the GaAs super cell, the symmetry lowering from cubic to tetragonal
take place and the $3d-t_{2g}$ states split further into a double degenerate
$d_{xz}$ and $d_{yz}$  situated at lower energies and the non-degenerate $d_{xy}$.
\begin{figure}[h]
\epsfig{figure=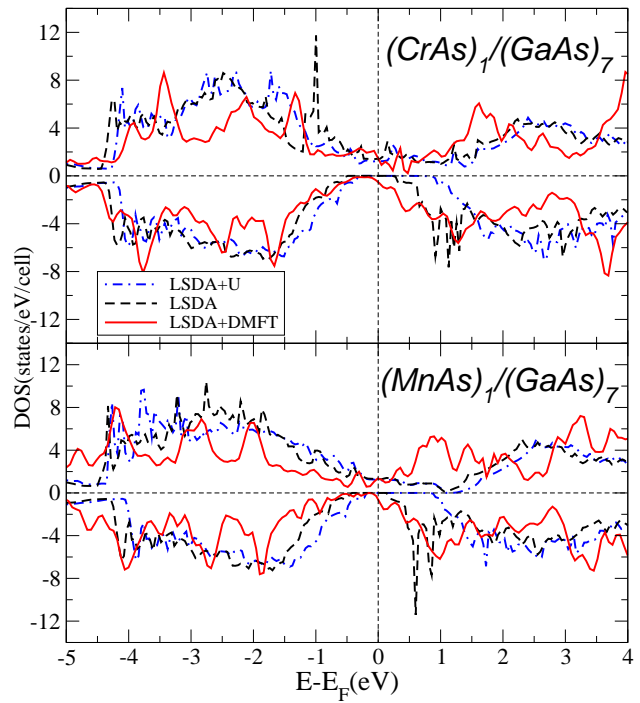,width=\linewidth}
\caption{(color online) Spin-resolved total DOS of (MAs)$_1$/(GaAs)$_7$, with M=Cr and Mn. 
The half-metallic character is evidenced by the simultaneous presence of a metallic 
majority-spin and an insulating (semiconducting) minority-spin DOS obtained within 
LSDA (dashed black-line) and LSDA+U (dot-dashed blue line) calculations. Dynamical 
correlations captured by DMFT (solid red-line) introduce many-body states within 
the gap that destroy half-metallicity.}
\label{Fig2_DOS}
\end{figure}
At higher energies the $d_{3z^2-1}$ and  $d_{x^2-y^2}$ are situated \cite{grif.64,im.fu.98}. 
Magnetic exchange interaction split the spin channels and contribute 
additionally to the crystal field in opening the minority spin gap.
In Fig.\ref{Fig2_DOS} we present spin resolved, total density of states (DOS), for the
DMH structures (MAs)$_{{\rm 1}}$/(GaAs)$_7$, M=Cr, Mn.
In both minority and majority spin channels at energy of 
about $-12eV$ (not shown in Figs.\ref{Fig2_DOS}) the As-$4s$ and Ga-$4s$ are present. 
In the energy range $-8$ to $-2eV$ As$-p$ and Ga$-p$ states are dominant characters in the 
$p-d$ bonding orbitals, while in the vicinity of the Fermi level in the energy 
range of $-2$ to $2eV$ the dominant $3d$ transition metal states are present
(see also orbtial projected DOS of Fig.\ref{Fig2l_DOS_rezolved}).  
At the LSDA level the occupied minority spin channel for both compounds 
is similar, because in this energy range the transition metal $d$-states are weakly 
present within the $p-d$ bonding orbital dominated by As and Ga-$p$ states. 
The Fermi level is situates for minority spin electrons within the gap, closer to the bottom 
of the conduction band. The values of the energy gaps are  
$\Delta _{Cr}^{LSDA} \cong 0.51eV$ and $\Delta _{Mn}^{LSDA} \cong 0.42eV$. The difference between the 
Fermi level and the bottom of the conduction band is somewhat larger for the DMH with 
Cr, in comparison with the value obtained in the Mn-based DMH. Given the fact 
that the top of minority-spin valence band is situated at similar energies in these 
heterostructures, the magnitude of the gap is therefore determined by the energy distance between 
the Fermi level and the bottom of conduction band.

A more detailed analysis of the density of states can be made on the base of the orbital projected
densities of states Fig. \ref{Fig2l_DOS_rezolved}. Here the LSDA and the LSDA+U results for the
correlated transition metal atoms are compared. 
\begin{figure}[h]
\epsfig{figure=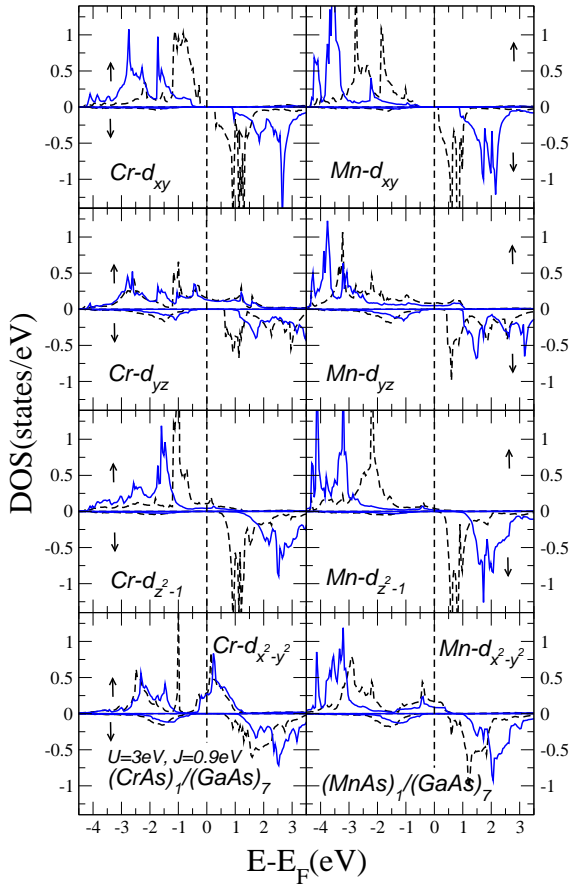,width=\linewidth}
\caption{(color online) Orbital projected spin polarized DOS for
(MAs)$_1$/(GaAs)$_7$, with M=Cr (left column) and Mn(right column). The
LSDA/LSDA+U results are shown with black-dashed/blue-solid lines.}
\label{Fig2l_DOS_rezolved}
\end{figure}
LSDA+U calculations were performed for different values of the average Coulomb
interaction $U=2,3,4$ and $5eV$ taking the same value for the exchange
parameter $J=0.9eV$. In Fig. \ref{Fig2l_DOS_rezolved} we show the results for U=3eV and J=0.9eV,
typical values for transition metals~\cite{im.fu.98,ar.im.04}.  
The left column represents the results for the compound with Cr, while the right column shows the
densities for the compound containing Mn. In each pannel the upper/lower part contains the results
for the majority/minority spins. Although the similarities of the orbital projected DOS are visible,
we note the general tendency of the LSDA+U namely to enlarge the minority spin gap
$\Delta _{Cr}^{LSDA+U} \cong 0.85eV$ and $\Delta _{Mn}^{LSDA+U} \cong 0.81eV$. An interesting
feature is that most of the spectral weight at the Fermi level is provided by the $3d_{x^2-y^2}$ orbital
for both cases of Cr and Mn, however the larger value is obtained with the Cr substitution. An almost
negligible spectral weight is obtained for $3d_{xy}$-orbitals.
As expected from this approach transition metal {\it d}-states are further pushed apart
from the Fermi level. In particular for the Cr heterostructure within the
majority spin channel the LSDA peaks (see Figs. \ref{Fig2_DOS},\ref{Fig2l_DOS_rezolved}) 
located at about -1eV are shifted to lower energies
at -1.5eV. The occupied majority spin Mn states
are within the bonding band at -3eV are also shifted to lower energies at -4eV.
Within the minority spin channel for all compounds  the group of states situated
around 1eV above $E_F$ are shifted accordingly to higher energies (see Fig. \ref{Fig2_DOS},
\ref{Fig2l_DOS_rezolved}).
For the other values of U and J a similar tendency of DOS was obtained for both
heterostructures.
\begin{figure}[h]
\epsfig{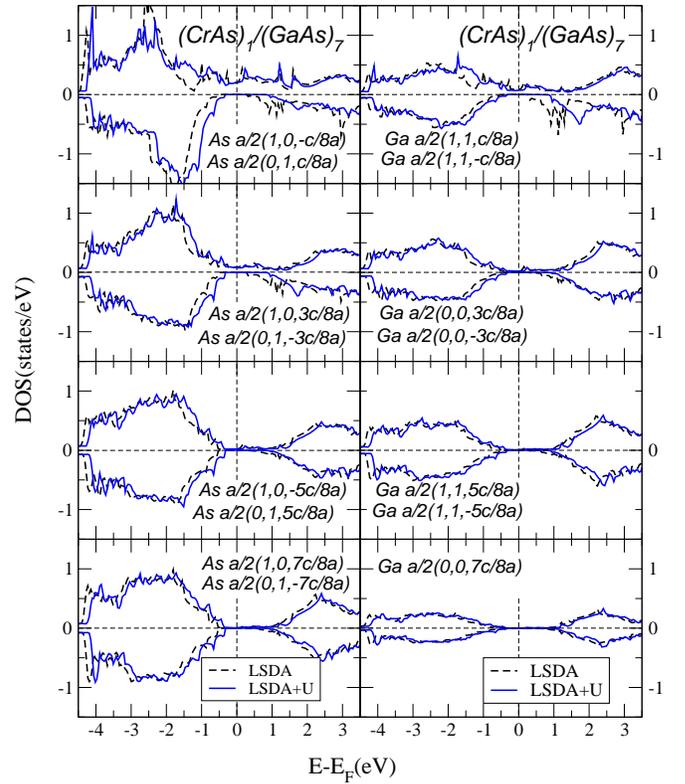}
\caption{(color online) LSDA (black-dashed) and LSDA+U (blue-solid)
As (left column), Ga(right column) spin polarized DOS for (CrAs)$_1$/(GaAs)$_7$. 
The positions in the unit cell are specified within the figures.
The U=3eV and J=0.9eV parameters were used in the LSDA+U calculation.}
\label{Fig2cras-ga}
\end{figure}
In the followings we plot the total As and Ga spin
resolved density of states for (MAs)$_1$/(GaAs)$_7$, M=Cr,Mn. Fig.\ref{Fig2cras-ga}
presents the results for Cr-$\delta$ doping. The left/right columns show the As/Ga-DOS 
for the different layers. As one can see in the majority spin channel the most 
important contribution around the Fermi level comes from the arsenic and gallium 
atoms situated at $As:a/2(1,0,c/8a), a/2(0,1,-c/8a)$ and $Ga:a/2(1,1,c/8a)$. These are in fact
the positions in the immediate vicinity of the transition metal CrAs-layer. 
The similar trend is visible for the Mn-$\delta$ doped heterostructure in Fig.\ref{Fig2mnas-ga}. 
In addition we note that as expected at the Fermi level the dominant
states are As-{\it p} states.
\begin{figure}[h]
\epsfig{figure=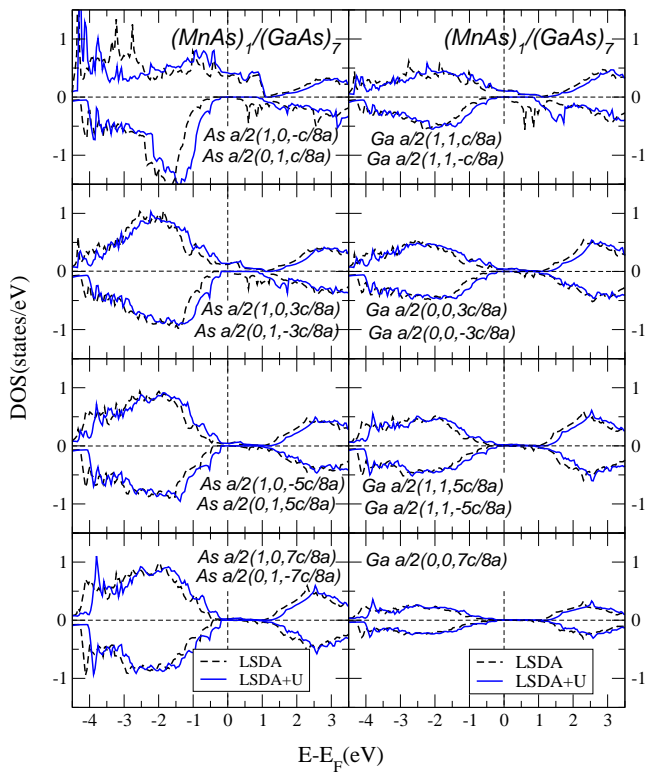,width=\linewidth}
\caption{(color online) LSDA (black-dashed) and LSDA+U (blue-solid)
As/Ga spin polarized DOS for (MnAs)$_1$/(GaAs)$_7$ are shown on the left/right
columns. The positions in the unit cell are specified within the figures.
The U=3eV and J=0.9eV parameters were used in the LSDA+U calculation.}
\label{Fig2mnas-ga}
\end{figure}
The general picture resulting from the extended analysis (by inspecting the atom and 
orbital resolved densities of states -Figs.\ref{Fig2l_DOS_rezolved}, \ref{Fig2cras-ga} 
and \ref{Fig2mnas-ga} ) of the LSDA and LSDA+U results is that the $\delta$-doped 
heterostructure presents a half-metallic behavior with a minority spin gap.

\begin{figure}[h]
\epsfig{figure=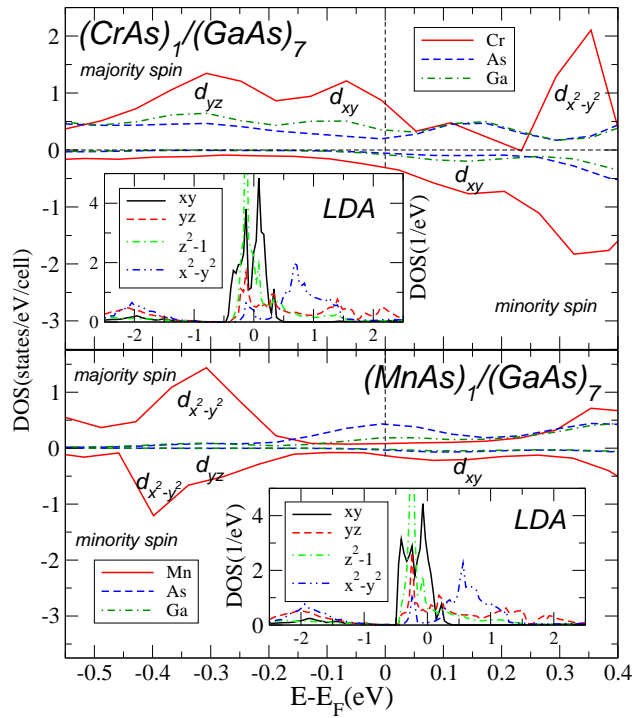,width=\linewidth}
\caption{(color online) LSDA+DMFT atom resolved spin polarized DOS for
(MAs)$_1$/(GaAs)$_7$, with M=Cr (upper panel) and Mn(lower panel).
The results presented were obtained for the following parameters:  $U_{Mn/Cr}=3eV$, $J_{Mn/Cr}=0.9eV$ and 
$T=50K$.
For the correlated atoms the main orbital contributions are
also shown (solid red line). Average total As (dashed blue)
and average total Ga (green dote dashed) contributions are
seen also. Above $E_F$, many-body induced states with
predominant $d_{xy}$ character determines the closure of the
half-metallic gap. Insets show the orbital resolved non-spin polarized Cr/Mn-LDA
density of states results.}
\label{Fig2a_DOS_rezolved}
\end{figure}

LSDA+DMFT calculations were performed for the same values of $U=3eV$ and $J=0.9eV$ 
and temperatures $T=30, 40, 50K$. In Fig. \ref{Fig2a_DOS_rezolved} we present the 
DMFT results for the atom resolved total density of states in a small energy window 
around $E_F$. Further we show the 3d-orbitals contribution into the main peaks
of Cr/Mn transition metals. The insets of Fig. \ref{Fig2a_DOS_rezolved} shows 
the results of non-magnetic calculations, from where we can roughly estimate 
the bandwidth which is of about 6eV. Based on the analysis of the values of the 
bandwidth, orbital occupations and the magnitude of U, one expects for the 
heterostructures in discussion, a behavior typical for a medium correlated metal.  
%
%In addition the LDA-orbitals occupation 
%can also be obtained: $n^{Cr/Mn}_{xy}\approx 0.9/1.5$, $n^{Cr}_{yz}\approx 0.8/0.9$,
%$n^{Cr/Mn}_{z^2-1}\approx 1.2/1.4$, $n^{Cr/Mn}_{zx}=n^{Cr/Mn}_{yz}\approx 0.8/0.9$, and 
%$n^{Cr/Mn}_{x^2-y^2}\approx 0.6/0.7$.  

As one can see  
the dynamic correlations captured by DMFT gave a completely different picture
for the orbital distributions around the Fermi level. In the case of Cr significant
majority spin density of states is obtained for the $3d_{xy}$-orbitals in contradiction
with the LSDA/+U results were these orbitals are shown to give a negligible contribution
see Fig. \ref{Fig2l_DOS_rezolved}.
Above $E_F+0.3eV$, in the unoccupied part the majority spin 3d electrons have a significant 
$3d_{x^2-y^2}$ and at higher energies (not shown) $3d_{z^2-1}$ character. In the minority spin channel from $E_F-0.5eV$ 
no significant spectral weight is present, while above $E_F$ the many body induced states 
have a predominant $3d_{xy}$ character. At higher energies a predominant $3d_{z^2-1}$ character  
is present.

In the case of Mn-$\delta$ doped structure, in the majority spin channel As-{\it p} is dominant 
at the Fermi level, the Mn-$3d_{x^2-y^2}$ contribution is seen around 0.3eV below the 
Fermi level. It is interesting to note that the position of this orbitals are not changed 
with respect to the LSDA picture. Contrary to the LSDA minority spin Mn-$3d_{x^2-y^2}$
have a larger spectral weight below $E_F$, while above $E_F$ $3d_{xy}$-states 
are seen to cross the Fermi level. The general picture obtained by including dynamic 
correlations is that the gap within the minority spin channel is closed and the 
half-metallicity disappears. For both Cr and Mn compounds closure of the gap is 
due to the presence of many-body states with $3d_{xy}$ character only seen in 
DMFT. 

\begin{figure}[h]
\epsfig{figure=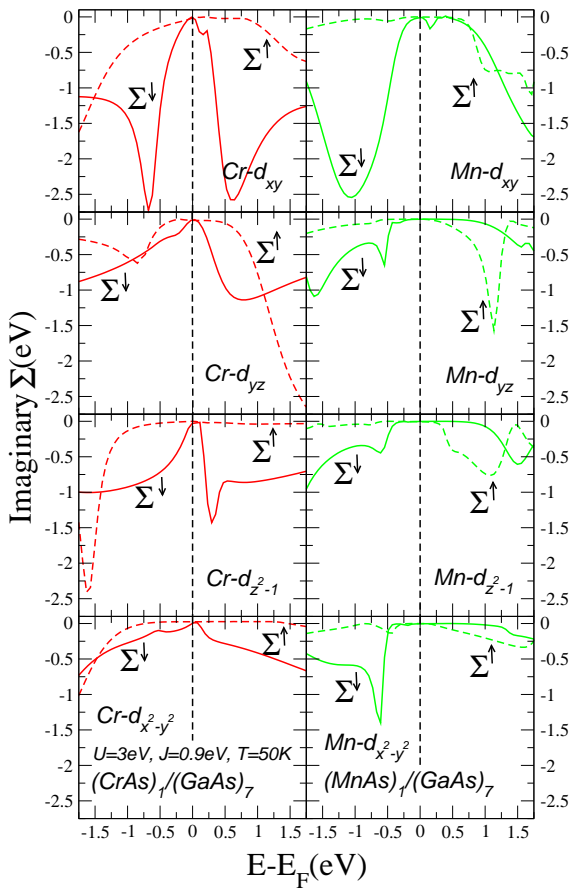,width=\linewidth}
\caption{(color online) Orbital resolved imaginary part of minority/majority  
spin self-energy for Cr (red solid/dashed) and Mn (green solid/dashed) lines 
computed for U$_{Mn/Cr}$=3eV and J=0.9eV and T=50K. The hump at around 0.2eV visible for
$d_{xy}$ orbital signals the departure from the expected Fermi liquid behavior.}
\label{Fig3_SIG}
\end{figure}

The analysis of the 
orbital resolved density of states can be correlated with the behavior of the 
self-energy seen in  Fig.\ref{Fig3_SIG}, 
where the atom and orbital resolved imaginary part of self-energy is shown.
The self-energies for $d_{zx}$-orbital were excluded since they are identical
due to the cell symmetry with the $d_{yz}$ ones.
We observe that the imaginary part of the self-energy has a rather symmetric energy
dependence around the Fermi level, with a normal Fermi-liquid-type behavior
$Im \Sigma^\uparrow (E)_{Cr/Mn} \propto (E-E_F)^2$. The minority spin
$Im \Sigma^\downarrow (E)_{Cr/Mn}$ shows a significant increase
just above the Fermi level which is more pronounced for
the $d_{xy}$ orbitals.
%
%Therefore NQP states within the minority spin gap of DOS are mainly determined 
%by the transition metal d$_{xy}$ orbitals.

\begin{figure}[h]
\epsfig{figure=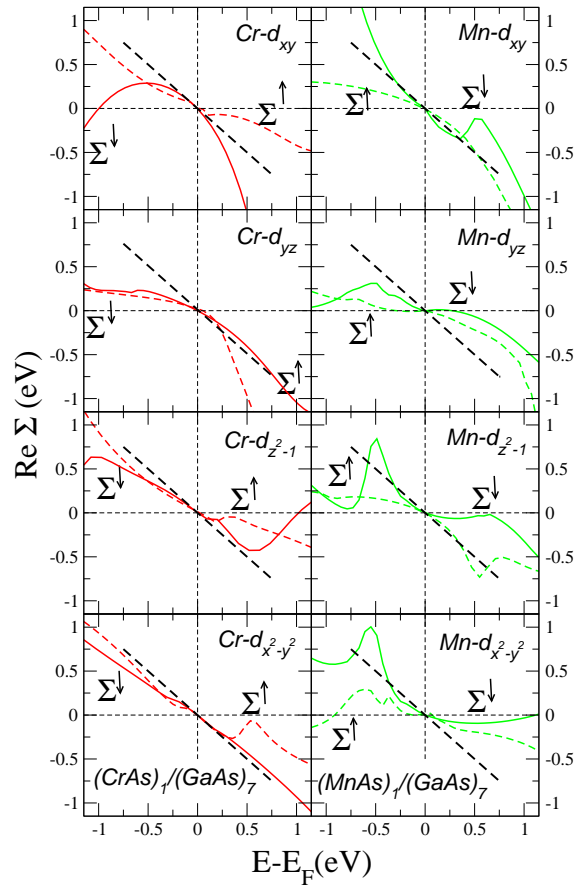,width=\linewidth}
\caption{(color online) Orbital resolved real part of minority/majority  
spin self-energy for Cr (red solid/dashed) and Mn (green solid/dashed) 
lines computed for U$_{Mn/Cr}$=3eV and J=0.9eV and T=50K. A solid black
line with a unitary negative slope is also shown.}  
\label{Fig4_rSIG}
\end{figure}
In Fig. \ref{Fig4_rSIG} we plot the real part of the orbital resolved majority and
minority self-energies. The left/right columns represent the results for Cr/Mn doped 
compounds. The figure also displays a line with a unitary negative slope as a guide 
for the analysis of the energy dependence of the real part of the self-energies around
$E_F$. As one can see for both spin 
directions, the self-energy has a negative slope $\partial \Sigma(\omega)/ \partial \omega$ 
at the Fermi energy, 
which confirms that the quasiparticle weight $Z=(1-\partial \Sigma(\omega)/ \partial \omega)^{-1}$
is reduced by correlations. For both Cr and Mn for majority spins
$\partial \Sigma(\omega)/ \partial \omega$ is clearly less than unity. 
Contrary for the minority Cr/Mn-$d_{xy}$ spins $\partial \Sigma_{xy}(\omega)/ \partial \omega$
is larger than unity above the Fermi level (within our approximation, we cannot determine 
with sufficient accuracy), suggesting the nonquasiparticle nature of
the minority-spin states within the gap. We have checked that the above behavior of the 
real and imaginary parts of the self-energy does not change significantly in the range 
of investigated temperatures. For smaller U values the slope is reduced slightly although, 
the same peculiar behavior is seen for the real and imaginary part of $d_{xy}$-orbital self-energy. 
Therefore we can conclude that the many-body induced NQP states within the minority spin 
gap situated just above the Fermi level, are mainly determined by the transition 
metal $d_{xy}$ orbitals. As the closure of the minority spin gap happens by the extension 
of NQP state tails from above, towards the Fermi level and in order to have a complete picture 
it is important to understand the effect of correlations upon the minority spin states 
situated below $E_F$. This can be studied by looking at the band edges within the minority 
spin channel. In particular we investigate the character of the electronic states below 
and above $E_F$ with or without including correlations. As discussed previously, above the
the Fermi level transition metal 3$d_{xy}$ states provide the main character that
developed many-body NQP states who's tail cross $E_F$. Below $E_F$ the dominant states
have As-{\it p} character. Therefore, we looked at contributions of As states originating from 
different As layers to the top of minority spin valence band. 
The As layers are grouped into four classes, each having the distance
$z_{As}= \frac a 2 \frac{jc}{8a}$, where $j=\pm1,\pm3,\pm5,\pm7$ with 
respect to the transition metal atom located in origin. The results of this 
analysis is presented in table \ref{tab-as}. 
\begin{table}[h]
\begin{tabular}{|l|cc|cc|}
\hline
& \multicolumn{2}{c|}{(CrAs)$_1$/(GaAs)$_7$} & \multicolumn{2}{c|}{(MnAs)$_1$/(GaAs)$_7$} \\
\cline{2-5}
           & $p_x$ & $p_y$ & $p_x$ & $p_y$ \\
           & $\%$  & $\%$  & $\%$  & $\%$  \\
\hline
As1:$\frac a 2 (0,1,\frac{c}{8a})$   & 10.14    & 16.66 & 10.11    & 17.23 \\
As1:$\frac a 2 (1,0,-\frac{c}{8a})$  &          &       &          &       \\
As2:$\frac a 2 (0,1,\frac{3c}{8a})$  & 27.57    & 22.22 & 27.03    & 20.43 \\
As2:$\frac a 2 (1,0,-\frac{3c}{8a})$ &          &       &          &       \\
As3:$\frac a 2 (0,1,\frac{5c}{8a})$  & 30.43    & 30.55 & 30.18    & 31.70 \\
As3:$\frac a 2 (1,0,-\frac{5c}{8a})$ &          &       &          &       \\
As4:$\frac a 2 (0,1,\frac{7c}{8a})$  & 32.88    & 30.55 & 32.66    & 30.64 \\
As4:$\frac a 2 (1,0,-\frac{7c}{8a})$ &          &       &          &       \\
\hline
\end{tabular}
\caption{Relative contribution (in percent) of the As-{\it p} states to the
valence band maxima from each As-layer obtained within the
LSDA calculation. Local correlation within the mean-field LSDA+U or DMFT
does not change significantly these values.}
\label{tab-as}
\end{table}
As the distance between the transition metal 
mono-layer and As mono-layer increases, the As-($p_{x}$, $p_{y}$) 
character continue to increase. In the same time, the difference between 
values from different mono-layers decreases. These results suggest that the 
As mono-layer situated closer to the transition metal has its electrons 
confined by the $p-d$ hybridization. At larger distances, hybridization 
decreases, and more $p$-character is available to form the top of the valence 
band. One can note that $p_x$ and $p_{y}$ contributions are 
different as a consequence of symmetry lowering present in the tetragonal 
geometry. These results are in agreement with previous calculations 
obtained using the macroscopic average of the total DFT potential \cite{sa.hi.01,sanv.03,ch.le.11}.

In the followings we comment the comparison of the results obtained within the 
three different approaches to the electronic structure calculations within the 
DFT. A well known problem of the LSDA and its relative GGA is the underestimation of the 
band gaps (for many semiconductors) in addition to the missing electronic
correlation effects. For the heterostructures in discussion we present results
obtained with different methods that consider correlations effects: a simplified mean field 
LSDA+U and a more complex LSDA+DMFT approach. The LSDA+U 
scheme seeks to correct the problems of LSDA by adding a repulsive potential +U, 
obtained from a mean field decoupling of the local Coulomb interaction. Its 
consequence is that the occupied/unoccupied 3d levels are shifted to lower/higher 
energies as seen in Figs. \ref{Fig2l_DOS_rezolved},\ref{Fig2cras-ga},\ref{Fig2mnas-ga}
The ground state picture resulting from both 
this methods is a half-metallic ferromagnetic state.

On contrary to the mean-field methods, DMFT captures correctly the dynamics at low
as well as high energies. This approach describes properly the quasiparticle formation around
Fermi level simultaneously with the lower and upper Hubbard bands at higher energies
\cite{ko.vo.04,ko.sa.06}. 
Dynamical correlations captured by DMFT change completely the physical picture of these compounds. 
For the majority spin states a slight spectral weight redistribution takes place, however this 
does not affect the value at the Fermi level which is similar to that obtained within LSDA
(see Fig. \ref{Fig2_DOS}). For minority spins, although the position of the top of the minority 
spin valence band is slightly changed including correlations, the As-{\it p} contributions 
from different As layers are similar in values to those obtained 
within the LSDA calculations, see Tab.~\ref{tab-as}. This demonstrates 
that the As-{\it p} character distribution within the top of the minority spin valence 
band is not sensitive to correlation effects.    
\begin{figure}[h]
\epsfig{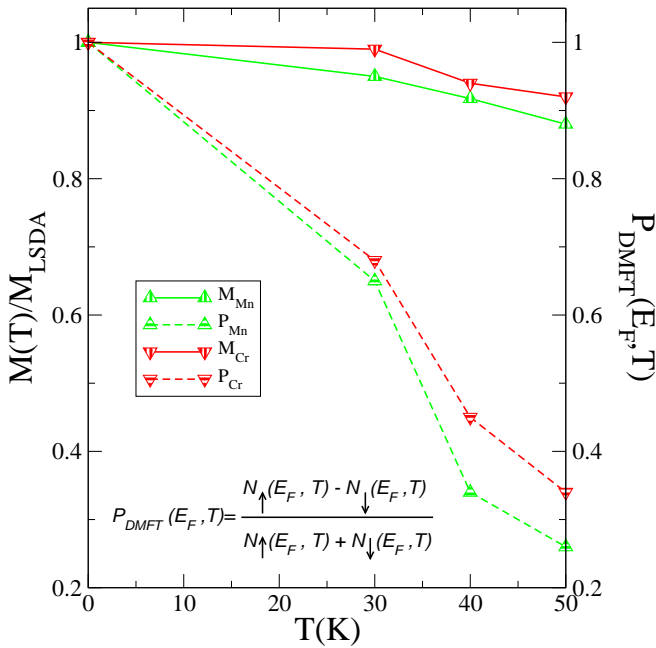}
\caption{(color online) LSDA+DMFT values for the finite temperature magnetization and spin polarization 
of (MAs)$_1$/(GaAs)$_7$, M=Cr, Mn compounds. The LSDA values are taken as T=0K results.}
\label{Fig4}
\end{figure}
The conclusion therefore is that the many-body interactions affects the half-metallic band structure
predominantly within the minority spin channel, above $E_F$, by introducing NQP states. 
The presence of these many-body states reduce significantly the ideal 
$100\%$ carrier spin polarization predicted by LSDA(+U). In order to discuss 
quantitatively the conduction electrons spin polarization we use the common definition 
\cite{mazi.99,ba.mu.07b,st.kr.08}: $P_{n}=(N_\uparrow(E_F)v^n_\uparrow(E_F)-N_\downarrow(E_F)v^n_\downarrow(E_F))/
(N_\uparrow(E_F)v^n_\uparrow(E_F)+N_\downarrow(E_F)v^n_\downarrow(E_F))$, where $N$ and $v$
are the density of states and the velocity at $E_F$. The values of $n$ determine what type of  
experimental measurements that can access the polarization. The $P_0$ corresponds to the 
spin-resolved photoemission experiments, while the higher orders corresponds to spin polarizations
measured with point contact Andreev reflection in the ballistic (n=1) or diffusive regimes (n=2)
\cite{mazi.99,ba.mu.07b}. In Fig. \ref{Fig4} we show the results for the polarization obtained 
using the simplified $P_{DMFT}(E_F,T)=P_{n=0}$, where the velocities for spin up and down are droped. 
One can see a clear distinction between the different temperature
behavior of magnetization and polarization, similar to many potential
half-metallic ferromagnets~\cite{ch.ka.03,ch.ar.06,ch.sa.08,ch.ar.09,ch.le.11}.

In table \ref{tab_mag_mom}  we present the values of the 
finite temperature magnetizations, the values of polarization
being easily readable from the lower panel of Fig. \ref{Fig4}. 
\begin{table}[h]
\begin{tabular}{|l|c|c|ccc|}
\hline
           & $M_{LSDA}$ & $M_{LSDA+U}$ & \multicolumn{3}{c|}{$M_{DMFT}$}   \\
\cline{4-6}
           & $\mu_B$    & $\mu_B$      & $30K$ & $40K$ &  $50K$     \\
\hline
{(CrAs)$_1$/(GaAs)$_7$} & 3.00 & 3.00         & 2.97    & 2.82  & 2.76         \\
{(MnAs)$_1$/(GaAs)$_7$} & 4.00 & 4.00         & 3.80    & 3.67  & 3.52         \\
\hline
\end{tabular}
\caption{Integer magnetic moments values were obtained within LSDA(+U).
The finite temperature DMFT magnetic moments demonstrate departure from the
half-metallic values.}
\label{tab_mag_mom}
\end{table}
For the Mn-$\delta$
doped GaAs the temperature dependence of remnant magnetization of samples that
consist of 100 period of GaAs(10 mono-layers)/MnAs(0.5 mono-layer) showed 
a quasi-linear dependence and a critical temperature of about $T_c \approx 50K$~\cite{ka.jo.00}.
More recent measurements on the Mn-$\delta$-doped DFH include polarized
neutron reflectometry and magnetometry performed on samples having
20, 50 and 100 mono-layer GaAs spacers in the temperature range from
30 to 40K~\cite{di.bo.04}. This analysis allowed to identify two types
of contribution to the magnetization of Mn-doped DMH: one type that has the
superlattice periodicity and the second one resulting from random clustering
effects suggested by the discontinuous nature of the quasi-two-dimensional
MnAs~\cite{di.bo.04}.
A direct comparison with these experimental results are
not possible for the present theoretical calculations, due to the fact that
the supercells considered here contain a considerably smaller number of layers.
Nevertheless certain qualitative agreements are obvious.
As one can see in Fig. \ref{Fig4} the present LSDA+DMFT calculation  
captures a qualitative quasi-linear temperature dependence of the reduced $M(T)/M_{LSDA}$ 
magnetization also for the smaller numbers of mono-layers in discussion, a behaviour that is
seen in the experiment. In addition the quasi-two-dimensional nature of the 
problem is reflected in the present calculations by the separation of electrons 
and holes among different layers (see Tab. \ref{tab-as} and the discussion below). 
A major difference is that for the current supercell geometry our estimate for 
the Curie temperature is around four times larger. 

Let us further comment upon the recent polarization measurements of spin carriers 
in Mn-$\delta$ doped GaAs compounds using hot 
electron photoluminescence (HPL)~\cite{sa.tr.08}. The hot-electron photoluminescent spectra
were interpreted based on the previous analysis performed in diluted magnetic 
semiconductors~\cite{sa.mo.05}. According to this analysis in Mn doped
GaAs in a wide range of Mn content the valence band holes predominantly occupy the 
Mn acceptor impurity band. It was suggested that the polarization of the valence band holes
is proportional to the sample magnetization~\cite{sa.mo.05,sa.tr.08}.
To interpret the data on digital Mn/GaAs heterostructures, a combined 
contribution from the $\delta$-doped and the unintentionally doped interlayer 
region was considered as a superposition of both contributions. It was shown that 
carriers experience a strong exchange interaction with the ferromagnetic
interfaces, which rapidly decease with increasing distance with respect to 
the ferromagnetic Mn-$\delta$ layer \cite{sa.tr.08}. 

As one can see the electronic correlations play an important role in the properties
of the defect-free DMH presented above. In the presence of dynamical many-body correlations 
the top of the valence band does not change significantly, however, 
NQP states appear in the minority spin channel, Fig.\ref{Fig2_DOS} and \ref{Fig2a_DOS_rezolved}, just 
above the Fermi level. At finite temperatures the tails of the NQP state 
cross the Fermi level and the half-metallicity is lost. In addition we have shown
that carriers spin polarization and magnetization follow a different temperature 
behavior. We expect that optically excited radiative recombination of electrons 
and holes within the minority spin channel would help to demonstrate the presence 
of the many-body effects in particular the existence of non-quasiparticle states 
in these materials. Hot-electron photoluminescence spectroscopy was recently used
to discuss hole spin polarization in diluted magnetic semiconductors \cite{sa.sa.09}.
The HPL circular polarization under circularly polarized excitation provides detailed 
information on spin-relaxation mechanism~\cite{sa.mo.05,sa.ra.08}, and allows a finite temperature characterization
of the HPL polarization~\cite{sa.sa.09}. We believe that such experiments would demonstrate a different 
finite temperature behavior of magnetization and polarization as qualitatively demonstrated 
in Fig. \ref{Fig4}.     

\section{Conclusions}
The technology based on digital alloy technique is expected to produce 
heterostructures with an enhanced $T_C$ because of locally confined high 
transition metal concentration, and a better control of the ferromagnetic 
properties by controlling the GaAs layer thickness. The expected enhancement 
of the Curie temperature has not yet been realized, instead, it was found 
that magnetic properties depend strongly on interlayer thicknesses~\cite{ka.jo.00,di.bo.04}. 
Our results of the electronic structure calculations based on LSDA predict 
that the defect free delta doped digital magnetic heterostructures are 
half-metals in agreement with previous calculations~\cite{sanv.03}. 
The same conclusion is given by computations that are using a mean field LSDA+U 
approach. On contrary including many-body correlations captured by DMFT the 
half-metallicity is lost. The computed finite temperature magnetization follows 
an almost linear temperature dependence similarly to the experimental measurements 
performed however for larger GaAs(10 mono-layers)/MnAs(0.5 mono-layer) 
heterostructures~\cite{ka.jo.00,di.bo.04}. In comparing the critical temperatures, our 
results overestimate by a factor of four the measured values. Certainly additional 
work is necessary in this direction in particular the extension of the theoretical 
approach to include dynamical correlations beyond locality. 
A particular importance in applications of DMH in spintronic devices is the 
effect of the ferromagnetic layers on the spin polarization of the carriers. 
Our results demonstrate that magnetization and polarization follows a different 
temperature dependence. We suggest that such effect might be captured 
by hot-electron photoluminescence spectroscopy. The quantitative analysis 
of the intensity of recombination radiation for transition between the 
electrons and holes within the minority spin channel in the presence 
of electronic correlations is in progress.

\section{Acknowledgements:}
L.C. and E.-V.M., acknowledge the support from the PN II ID\_672/2008 grant and the FWF
projects no. P19630-N16, P18551-N16, P21289 and by the cooperation project NAWI Graz Grant
No. F-NW-515-GASS. L.V. acknowledges the financial support from the Swedish Research 
Council and the Hungarian Scientific Research Fund (research projects OTKA 84078).

\bibliography{references_database}
\end{document}